\documentclass[lettersize,journal]{IEEEtran}

\usepackage{amsmath,amsfonts}
\usepackage{algorithm}
\usepackage{algpseudocode}
\usepackage{graphicx}
\usepackage{booktabs}  % 用于绘制表格的横线，使表格更美观
\usepackage{multirow}  % 引入 multirow 宏包，支持多行单元格合并
\usepackage{subcaption}  % 替代subfigure，功能更完善
\usepackage{url}
\usepackage{verbatim}
\usepackage{array}
\usepackage{stfloats}
\usepackage{hyperref} % 关键！实现点击跳转
\usepackage[utf8]{inputenc}
\usepackage{xcolor}
\def\BibTeX{{\rm B\kern-.05em{\sc i\kern-.025em b}\kern-.08em
    T\kern-.1667em\lower.7ex\hbox{E}\kern-.125emX}}
\usepackage{balance}

\begin{document}

\title{MGDiff: Multi-Interest Sequence Recommendation with Masking GNN-Guided Diffusion}
\author{IEEE Publication Technology Department
\thanks{Manuscript created October, 2020; This work was developed by the IEEE Publication Technology Department. This work is distributed under the \LaTeX \ Project Public License (LPPL) ( http://www.latex-project.org/ ) version 1.3. A copy of the LPPL, version 1.3, is included in the base \LaTeX \ documentation of all distributions of \LaTeX \ released 2003/12/01 or later. The opinions expressed here are entirely that of the author. No warranty is expressed or implied. User assumes all risk.}}

\author{Wenjing Xiao, Hao Ding

% \author{Wenjing Xiao, Hao Ding, Zhe Zhang, Min Liu, Jonathan M. Garibaldi,~\IEEEmembership{Fellow}, Yan Zhang,~\IEEEmembership{Fellow}
        % <-this % stops a space
% \thanks{W. Xiao and H. Ding are with the School of Computer, Electronics and Information, Guangxi University, Nanning 530004, China, and also with the Guangxi Key Laboratory of Multimedia Communications and Network Technology, Guangxi 530004, China, E-mail: wenjingx@gxu.edu.cn, haoding@st.gxu.edu.cn.
% }
% \thanks{Z. Zhang is with the School of Electrical Engineering,Guangxi University, Guangxi 530004, China, qyzz@hnu.edu.cn.}
% \thanks{M. Liu is with the School of Artificial Intelligence and Robotics and National Engineering Research Center of Robot Visual Perception and Control Technology, Hunan University, Changsha, Hunan, China. Email:liu\_min@hnu.edu.cn.}
% \thanks{J. Garibaldi is with the School of Computer Science, University of Nottingham, NG8 1BB Nottingham, U.K. E-mail: jmg@cs.nott.ac.uk}
% \thanks{Y. Zhang is with the School of Information and Communication Engineering, University of Electronic Science and Technology of China, Chengdu 611731, China, E-mail: yanzhang@uestc.edu.cn.}
% % \thanks{Corresponding authors: M. Chen.}
}

% \markboth{IEEE Transactions on Neural Networks and Learning Systems}%
\markboth{Artificial Intelligence}%
{How to Use the IEEEtran \LaTeX \ Templates}

\maketitle
\begin{abstract}
Leveraging their ability to model data distributions and generate high-quality items, diffusion models (DMs) have demonstrated significant advantages in sequence recommendation. However, existing DM-based methods face challenges in recommendation quality and training efficiency, primarily due to: (1) Semantic distortion of the guidance information. User interaction histories are highly sparse and random. Modeling directly on raw signals tends to overfit superficial co-occurrence and introduce extraneous information into the guidance, thereby distorting semantics and degrading generation precision in diffusion models. (2) Endogenous generation bias. The probability-driven denoising process is prone to mode collapse towards high-density popular items. Consequently, popularity signals hijack the generative trajectory, overshadowing true user interests and leading to homogenized recommendations. To address these issues, we propose a novel Multi-Interest Sequence Recommendation Framework with \underline{M}asking \underline{G}NN-Guided \underline{Diff}usion Model (MGDiff), designed to generate accurate, bias-free user interest information during the diffusion process.
% First, we propose a semantics-enhanced Dual-layer Semantic Guidance (DSG) framework. Specifically, the DSG employs a semantic–interest decoupling mechanism that extracts structured latent semantics through a Weight-adaptive Masking Graph Neural Network (GNN) and achieves multi-dimensional, fine-grained decomposition of user interests via a dynamically routed Mixture-of-Experts (MoE). This collaboration produces structured semantic guidance information, enabling the DMs to achieve more precise generation.
First, we propose a semantics-enhanced Dual-layer Semantic Guidance (DSG) framework, which decomposes guidance into two synergistic stages: extracting latent item semantics and decoupling multidimensional user intent. We design a Weight-adaptive Masking Graph Neural Network reconstructs missing links to uncover deep item relationships beyond superficial co-occurrence, while a Dynamic Multi-Expert Network projects user preferences into distinct semantic subspaces to suppress irrelevant interference. This hierarchical design yields structured guidance that significantly improves the generation accuracy of diffusion models. Second, We propose a Popularity-Aware Guidance (PAG) mechanism that performs spatial geometric adjustments on the outputs of diffusion models: by using item popularity as a differentiable adjustment signal to recalibrate similarity metrics, we enable DMs to generate diverse recommendations free from popularity bias.
Finally, we compare MGDiff with multiple baseline models across four widely used datasets, demonstrating its superior performance and validating its effectiveness. 
% Code will be made available upon publication.
\end{abstract}

\begin{IEEEkeywords}
Recommendation Systems, Graph Mining, Social Network Analysis, Diffusion Model.
\end{IEEEkeywords}

\section{Introduction}
Sequence recommendation (SR) is a fundamental task in personalized recommendation systems. It leverages users' historical interaction sequences to capture their interests and predict their next items of interest~\cite{LLMEmb:14}. With advances in deep generative models~\cite{Deep:21,dong2024accelerating:23,liu2024milpstudio:24,TNNLsllm3}, diffusion models (DMs) have been gradually incorporated into sequence recommendation due to their superior generation quality and training stability.
Currently, diffusion model–based approaches mainly focus on modeling the latent distribution of user interests. Prior works commonly introduce noise to the target item and iteratively perform denoising guided by user interaction sequences, aiming to generate recommendations that are highly consistent with user preferences~\cite{DiffkG:16,Diffrec:10,DiQDif:1,TNNLS5G-Diff5,TNNLSdiffcl}. These approaches have demonstrated potential in enhancing 
quality of generated content.

% \begin{figure}[t]
%     \includegraphics[width=0.5\textwidth, height=0.3\textheight, keepaspectratio]{Figure/illu.drawio.pdf} % 图片宽度设为页面宽度的80%
%     \caption{Illustrations of (a) there are wrong relations / representation among items, and (b) user's true interests are often drowned out by the prevalence of popular items during model training, creating a bias in the system.}
%     \label{fig:1}
% \end{figure}

Despite these advancements, existing methods based on DMs still suffer from the following two primary limitations. 
(1)~\textit{Semantic distortion of the guidance information}. \textbf{Firstly}, the core objective of guidance information is to encode user interests based on their historical interaction sequences, thereby providing effective personalized conditions for the subsequent item generation process and improving recommendation accuracy~\cite{DDPM:1}.
However, existing methods, such as the attention-based guidance mechanism proposed by Li et al.~\cite{DimRec} and the cluster-based quantization guidance by Mao et al.\cite{DiQDif:1} are fundamentally constrained by the high sparsity and inherent randomness of raw interaction data. 
% As illustrated in Fig~\ref{fig:1}a, this intrinsic limitation causes models to overfit spurious correlations within the data. 需要恢复
For instance, a model might misidentify incidental co-purchases (e.g., beverages and electronics) as genuine user preferences, thereby forcing itself to over-learn these pseudo-relationships between unrelated items. In sparse and stochastic interaction environments, such noise is further amplified, resulting in severely distorted guidance signals that ultimately mislead the diffusion generation process toward trajectories deviating from the true semantics. Therefore, it is imperative to delve into the latent semantic structures underlying user-item interactions. This necessitates that the model moves beyond the mechanical bundling of co-occurrence signals or superficial pattern recognition, aiming instead to comprehend the inherent logical relationships between items.
(2)~\textit{Endogenous generation bias.} \textbf{Secondly}, our statistics show that in the \textit{Amazon Beauty dataset}, user sequences with cold items as target items account for approximately 47\%. 
% As shown in Fig.~\ref{fig:1}b, current methods tend to reduce the exposure of long-tail products, making it difficult to recommend them effectively and satisfy diverse user demands. 需要恢复
The root cause of this issue lies in the dominance of popular items in the training data. Specifically, in the context of diffusion models, their fundamental mechanism involves estimating and progressively removing the added noise to generate items. However, this denoising process is highly susceptible to mode collapse and generating overly similar outputs when data bias is present~\cite{DDPM3,DDPM5}. Consequently, the model overfits the high-frequency co-occurrence patterns of popular items~\cite{bias:1,bias:2}, learning item representations that overemphasize popularity signals rather than adequately capturing intrinsic item attributes (such as type, style). As a result, in downstream recommendation tasks, the model tends to over-recommend popular items, leading to repeated exposure. This not only reduces the diversity of recommendations but also severely limits the exposure opportunities for long-tail items, thereby exacerbating the Matthew effect in recommendation systems.

To address the aforementioned challenges, we propose MGDiff,  which effectively captures user interests and generates diverse recommendations.
Specifically, to overcome the first limitation, we propose a Dual-layer Semantic Guidance (DSG) framework. The core of DSG lies in decomposing the semantic guidance process into two progressively interlinked layers: the lower layer is responsible for extracting latent logical relationships among items, thereby providing the upper layer with structured semantic representations stripped of co-occurrence noise; the upper layer then performs multidimensional decoupling of user intent based on these representations, ultimately generating precise guidance signals for diffusion models. Specifically, to capture the latent semantic relationships between objects (that is, their intrinsic logical structure), we designed a Weight-adaptive Masking Graph Neural Network (GNN) to characterize the latent transfer dependencies between objects. By introducing a masking mechanism, the process of learning object relationships avoids simply replicating surface-level co-occurrence signals. Instead, it reconstructs missing connections to uncover deep, intrinsic semantics from higher-order neighborhood topologies. At the same time, the adaptive weighting strategy enables the model to progressively capture more complex interaction patterns, ultimately learning robust object representations capable of semantic expression. Based on the item representations derived from the former, we constructed a Dynamic-Multi Expert Network to identify user intent. This module decomposes user preferences into multiple semantic subspaces to mitigate the interference of irrelevant information during fitting. During training, the routing mechanism gradually transitions from an exploratory mode to a specialized mode, enabling items to first explore diverse semantic associations and then focus on the most relevant interest components. The former is dedicated to stripping away noise from co-occurrence signals and recovering the intrinsic structural semantics among items; the latter further decomposes user preferences into subspaces on this foundation, thereby preventing irrelevant information from confounding the interest fitting process. In this way, DSG provides precise semantic guidance for DMs.
% To address the second limitation, we propose a Popularity-Aware Guidance (PAG) strategy that converts item popularity into a differentiable similarity adjustment signal and employs a contrastive loss function to enhance the discriminability between items, effectively mitigating popularity bias. This strategy allows the recommendation system to balance global popularity trends and individual preferences, providing a more comprehensive modeling perspective for capturing user interests.
To address the second limitation, we propose a Popularity-Aware Guidance (PAG) strategy. Specifically, we transform item popularity into a differentiable similarity modulation signal, dynamically reweighting the learning contributions of different items during training to suppress the dominance of popular items and enhance the gradient signals of long-tail items. Through this mechanism, PAG adaptively reshapes the embedding space structure, mitigating the issue of the diffusion model collapsing toward similar popular items during denoising, thereby enabling the model to strike a more reasonable balance between popular trends and individual preferences.

Specifically the contributions we made:
\begin{itemize} 
\item To address the semantic distortion of guidance information, we propose a Weight-adaptive Masking GNN combined with a dynamic routing multi-interest Mixture-of-Experts  network. This framework effectively extracts latent structures to generate precise and stable guidance for the diffusion model. 
\item We propose a Popularity-Aware Guidance  strategy to mitigate endogenous generation bias. This mechanism enhances the discriminability of items, effectively solving the issues of low exposure for long-tail items and homogenized recommendations. 
\item We conduct extensive experiments on four real-world datasets, demonstrating that our proposed approach consistently outperforms state-of-the-art baseline models. 
\end{itemize}

\section{Method}

% \subsection{Problem Definition}
Let $\mathcal{I}$ denote the set of items, and let $Q = [v_1, v_2,\ldots,v_{L-1}]$ represent the user's interaction sequence, where $v_i$ is the $i$-th item in chronological order. The ground-truth next item that the user will interact with is denoted as $v_L$. Previous studies~\cite{DiQDif:1,LLMEmb:14,Diffrec:10} typically use an encoder to generate high-dimensional embedding vectors for the items, that is, the embedding representation of the user's sequence $Q$ is $X = [x_1, x_2, x_3, \ldots,x_{L-1}]$, where $x_i \in \mathbb{R}^d$, with $d$ representing the embedding dimension. By learning the relationships between these high-dimensional embeddings, the user representation is generated, which is then used to predict the next item $v_L$ that best matches the user's interests~\cite{DiQDif:1,Diffrec:10}.

\subsection{Overall Framework}

We propose the Dual-Layer Semantic Guidance framework, which models semantic guidance learning as two hierarchical and closely interrelated stages. We design a Weight-adaptive Masking Graph Neural Network, which employs a masking mechanism to enhance the learning of item embeddings. The resulting item embeddings $G$ are fused with the user's initial interaction sequence encoding $X$ to form a refined user embedding vector $H$, which preserves both the original item features and the relational information among items. We further design a dynamic routing-based Mixture of Experts  module, which disentangles the user's representation $H$ containing item transition relationships, into multi faceted user interests. This provides accurate and robust guidance to the DM. DSG explicitly separates item relationship modeling from user interest decomposition and enables synergistic effects between the two: the structured representation from the first stage provides a semantic foundation for multi-interest decoupling in the second stage, while the multi-interest representation from the second stage, in turn, enhances the diffusion model’s understanding of user preferences, thereby achieving progressive bidirectional gains from item association to interest decomposition.
Meanwhile, the PAG module utilizes item popularity information to calibrate the similarity matrix derived from the diffusion model's final recommendation vector. It employs a contrastive loss to ensure that the model remains sensitive not only to popular items but also to less popular ones, thereby promoting recommendation diversity.
During training, MGDiff introduces Gaussian noise to the ground-truth next target item in the forward process and enhances conditional guidance using the refined embeddings from DSG. The denoising model is then trained to reconstruct the original items under this reinforced guidance, while simultaneously leveraging PAG to improve performance on non-popular items. During inference, the model starts from pure Gaussian noise and iteratively refines the prediction of the next item, guided by the embeddings provided by DSG.

\subsection{Dual-layer Semantic Guidance}
% In recommendation systems, users' browsing and purchasing behaviors generate rich item transition information. To extract accurate users' true interests from sparse and noisy interaction data, we propose Precise Interest Capture (PIC) module. PIC employs a masked GNN to capture the intrinsic connections between items for item embedding representation, while simultaneously leveraging a dynamic multi-interest expert network to achieve more fine-grained and semantically explicit representations of user interests. In the following, we elaborate on PIC from two perspectives: item embedding generation and user interest extraction.
% However, due to the sparsity and randomness of user interaction data, generating accurate guidance information for diffusion models remains a significant challenge. The DSG module addresses this issue by employing a masked GNN to model intrinsic item relationships and a dynamic multi-interest expert network to derive fine-grained and semantically explicit user interest representations.
% The following sections detail the design of DSG.
The DSG module consists of two core components: a weight-adaptive masking graph neural network and a multi-interest expert network. These two components work collaboratively, and their details will be elaborated in the following sections.

\textbf{Weight-adaptive Masking GNN.}
We construct a graph $\mathcal{G}=(\mathcal{I},\mathcal{E},\mathcal{W})$ based on the user behavior sequences in the training set, where the item set $\mathcal{I}$ serves as the node set. $\mathcal{E}$ is the edge set, and each edge connects two adjacent items in a user interaction sequence. By traversing all user sequences, we establish edge weights $w(i,j)\in \mathcal{W}$, where $w(i,j)$ denotes the number of transitions from item $i$ to item $j$. To capture deeper semantic relationships among items, we introduce a weight-adaptive masking mechanism. This strategy progressively increases the task difficulty by adhering to the principles of curriculum learning, thereby fostering more robust representation learning. This approach is motivated by the observation that high-weight edges, often connecting popular items within dense neighborhoods, are relatively easy to reconstruct even when masked. In contrast, low-weight edges, typically associated with long-tail items in sparse regions, present a more significant reconstruction challenge. 

Driven by this insight, our weight-adaptive strategy begins by preferentially masking the easily predictable high-weight edges with a high probability. As training progresses, it gradually shifts towards a uniform masking strategy, thereby increasingly exposing the model to the more challenging low-weight edges. Specifically, we compute the sampling probability for each edge using a linear decay schedule, ensuring a smooth transition from a biased to a uniform sampling distribution. The masking probability for the $i$-th edge $e_i$ is formulated as:
\begin{equation}
\tilde{p_i} = \min\left(1, \frac{w_i^\alpha \cdot r \cdot N}{\sum_{j=1}^N w_j^\alpha}\right),
\label{10}
\end{equation}
where \( w_i \) represents the weight of the \( i \)-th edge, \( r \) represents the sampling ratio, and \( N \) is the total number of edges. \( \alpha = \max\left(0, 1 - \frac{2 \cdot t}{T}\right) \), where $t$ is the current epoch and $T$ is the total number of epochs. The set of masked edges $\mathcal{E}_{\text{mask}}$ is subsequently generated via Bernoulli sampling based on these probabilities:
\begin{equation}
\mathcal{E}_{\text{mask}} = \left\{ e_i \,\Big|\, m_i = 1, \text{ and } m_i \sim \text{Bernoulli}(\tilde{p_i}) \right\}.
\label{11}
\end{equation}

During training, the masked edges are removed from the original graph $\mathcal{G}$ to yield a corrupted graph $\mathcal{G'}$. Let $\mathbf{H}^{(l)}$ denote the node representation matrix at the $l$-th layer, with $\mathbf{H}^{(0)} = A$ being the initial item feature matrix. The layer-wise propagation rule of our GNN encoder is then defined as:
\begin{equation}
{H}^{(l+1)} = \sigma \left( \mathrm{LayerNorm} \left( \tilde{{A}} {H}^{(l)} {W}^{(l)} + \mathbf{b}^{(l)} \right) \right),
\label{12}
\end{equation}
where $\tilde{{A}}$ is the adjacency matrix of the masked graph $\mathcal{G'}$, ${W}^{(l)}$ and ${b}^{(l)}$ are the trainable weight matrix and bias vector for the $l$-th layer, respectively, $\mathrm{LayerNorm}(\cdot)$ denotes layer normalization, and $\sigma(\cdot)$ is the ReLU activation function. The final output embedding matrix $G$ is obtained after $L$ layers of propagation, $G = {H}^{(L)}$.

Then, we concatenate the embeddings of the items connected by the masked edges, generated by the GCN, and use a multi-layer perceptron to predict whether the edge exists. 
\begin{figure}[t]
   \includegraphics[width=0.5\textwidth, height=0.35\textheight, keepaspectratio]{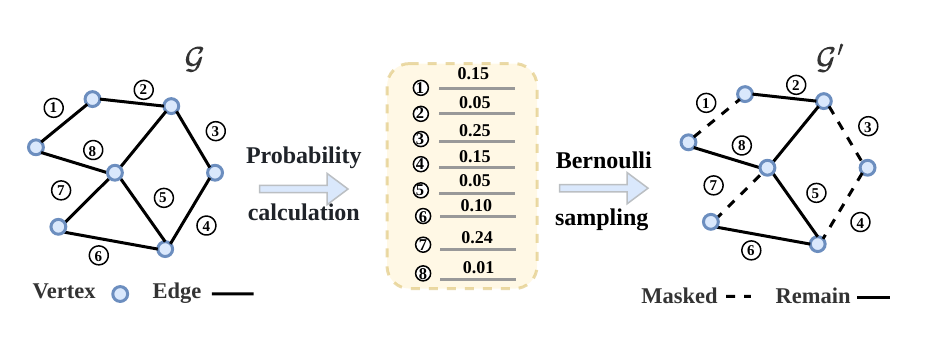} % 图片宽度设为页面宽度的80%
    \caption{Weight-adaptive Masking GNN.}
    \label{fig:3}
\end{figure}
For the pair of nodes $(i, j)$, the probability that an edge exists between them is calculated as follows:
\begin{equation}
s_{ij} = \sigma \bigl( \text{MLP} \bigl( [g_i \| g_j] \bigr) \bigr),
\label{13} %
\end{equation}
where $\sigma$ denotes the Sigmoid function, $[g_i | g_j]$ denotes the concatenation of node pair embeddings.
The process is optimized using a contrastive loss, where positive samples are the masked edges, and negative samples are randomly sampled from the non-existing edges in the graph $\mathcal{G}$. We can calculate the  loss $\mathcal{L}_{g}$ as: 
\begin{equation}
\begin{split}
\mathcal{L}_{g} &= -\frac{1}{|\mathcal{E}_{\text{mask}}|} \sum_{(i,j) \in \mathcal{E}_{\text{mask}}} \log(s_{ij} + \epsilon) \\
&- \frac{1}{|\mathcal{E}_{\text{neg}}|} \sum_{(u,v) \in \mathcal{E}_{\text{neg}}} \log(1-s_{uv} + \epsilon),
\end{split}
\label{14}
\end{equation}
where, $s_{ij}$ and $s_{uv}$ represent the prediction scores of the MLP decoder for the positive edge of the sample $(i,j)$ and the negative edge of the sample $(u,v)$, respectively. ${|\mathcal{E}_{\text{mask}}|}$ and ${|\mathcal{E}_{\text{neg}}|}$ represent the number of positive samples and negative samples, respectively.
Unlike traditional approaches that rely on explicit observations, our masking strategy serves as a bottleneck that prevents node $i$ from exploiting the direct shortcut to node $j$. Instead, it compels the model to infer the potential relationship by aggregating information from multi-hop neighbors. This mechanism discourages the mere memorization of superficial co-occurrence patterns and encourages the exploration of intrinsic semantic correlations, effectively revealing the underlying rationale behind user transitions.

% By drawing on the concept of human learning, a dynamic routing mechanism was designed to achieve a more granular and semantically explicit modeling of user interests.
% Because the traditional MoE (Mixture of Experts) architecture adopts hard routing (such as \text{Top-}$K$ selection), this results in non-differentiable operations. To address this,

\begin{figure}[t]
    \includegraphics[width=0.5\textwidth, height=0.5\textheight, keepaspectratio]{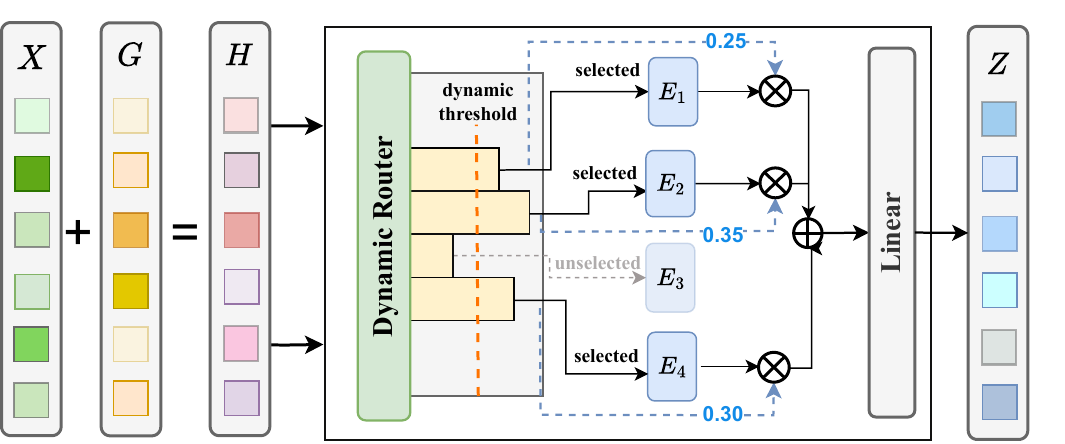} % 图片宽度设为页面宽度的80%
    \caption{Dynamic Multi-Experts Network}
    \label{fig:4}
\end{figure}

\textbf{Dynamic-Multi Expert Network.} As illustrated in Fig.~\ref{fig:4}, to reduce reliance on irrelevant relationships in user sequence analysis and generate more robust guidance information, we propose a sequence parsing method based on a dynamic mixture-of-experts network. Specifically, the input to the MoE is made up of two parts: the high-dimensional user embeddings $X = [x_1, x_2, x_3, \ldots,x_{L-1}]$ generated by the encoder, and the embeddings of elements $G = [g_1, g_2, g_3, \ldots,g_{L-1}]$ generated by the Weight-adaptive Masking GNN. We fuse these representations through element-wise summation to construct an integrated sequence representation:
$H = [h_1, h_2, \ldots, h_{L-1}] \quad \text{where} \quad h_i = x_i + g_i$.
We first use a gated network to generate expert weights logits $\boldsymbol{\alpha}_i \in {R}^k$, which are calculated as follows:
\begin{equation}
\alpha_i ={W}_2 \cdot \text{ReLU}({W}_1 \cdot {h}_i).
\label{15} %
\end{equation}
To obtain a differentiable approximation of discrete choice behavior, we use the Gumbel-Softmax technique to perturb and normalize $\boldsymbol{\alpha}_i$, yielding the soft expert allocation probability ${p}_i \in {R}^K$:
\begin{equation}
p_{i,k} = \frac{\exp\left( \frac{ \alpha_{i,k} + m_{i,k}}{\tau} \right)}{\sum_{j=1}^{K} \exp\left( \frac{\alpha_{i,j} + m_{i,j}}{\tau} \right)},
\label{16} %
\end{equation}
where, $\alpha_{i,k}$ represents the original matching score between the $i$-th token and the $k$-th expert, $m_{i,k} \sim \text{Gumbel}(0, 1)$ represents the noise term sampled from the Gumbel distribution, and $\tau$ is the temperature parameter used to regulate the discreteness of the sampling.
To achieve progressive expert specialization during training, we employ a dual annealing strategy. The temperature $\tau$ gradually decreases to sharpen expert assignments:
\begin{equation}
\tau = \max \left( \tau_{\text{final}},\  \tau_{\text{initial}} \cdot \exp(-\beta \cdot t) \right),
\label{17}
\end{equation}
where $\tau_{\text{initial}}$ represents the initial temperature, $\tau_{\text{final}} $ denotes the \textbf{minimum} temperature, $\beta$ controls the decay rate, and $t$ represents the training step. As $\tau(t)$ decreases, the Gumbel-Softmax distribution becomes sharper, concentrating probability mass on fewer experts. Simultaneously, we introduce a dynamic threshold $\theta(t)$ that gradually increases to promote sparser expert activation. The threshold is updated according to:
\begin{equation}
\theta(t) = \min \left( \theta_{\text{final}},\  \theta_{\text{initial}} \cdot \exp(\gamma \cdot t) \right),
\label{18}
\end{equation}
where $\theta_{\text{initial}} $ denotes the initial routing threshold, $\theta_{\text{final}}$ specifies the maximum threshold during training, $\gamma$ controls the growth rate, and $t$ represents the training step. As training progresses, $\theta(t)$ increases toward $\theta_{\text{final}}$, thereby restricting expert activation to only those with sufficiently large assignment probabilities and enforcing progressively stronger sparsity in the routing process.
The sparse routing mechanism only activates experts whose allocation probability exceeds the current threshold:
% \begin{equation}
% \mathbf{z}_i = \sum_{k=1}^{K} \mathbb{I}(p_{i,k} > \theta(t)) \cdot p_{i,k} \cdot \text{Expert}_k(\mathbf{h}_i),
% \label{19}
% \end{equation}
% where $\mathbb{I}(\cdot)$ is the indicator function, and $\text{Expert}_k(\cdot)$ represents the $k$-th expert network.
\begin{equation}
\begin{aligned}
\tilde{m}_{i,k} &= \text{sg}\bigl(\mathbb{I}(p_{i,k} > \theta(t))\bigr) + p_{i,k} - \text{sg}(p_{i,k}), \\
\mathbf{z}_i &= \sum_{k=1}^{K} \tilde{m}_{i,k} \cdot \text{Expert}_k(\mathbf{h}_i),
\end{aligned}
\label{19}
\end{equation}
where $\text{sg}(\cdot)$ denotes the stop-gradient operator, satisfying $\text{sg}(x)=x$ in the forward pass and $\frac{\partial \text{sg}(x)}{\partial x}=0$ in the backward pass. This formulation preserves the hard indicator $\mathbb{I}(p_{i,k} > \theta(t))$ for sparse expert activation in the forward pass, while replacing its gradient with that of $p_{i,k}$ during backpropagation. As a result, the routing mechanism supports end-to-end training via a straight-through gradient approximation, while maintaining computational efficiency.

This dual annealing strategy creates a synergistic effect: decreasing temperature sharpens the expert probability distribution, while increasing threshold raises the bar for expert selection. Consequently, the number of activated experts per token monotonically decreases throughout training, achieving the desired progressive specialization. We can obtain the user's final  embedding: $\mathcal{Z} = [z_1, z_2, \ldots, z_{L-1}]$.
We combine the user's initial  embeddings  $X$ with the embeddings generated by DSG to obtain the final precise guidance information for the DMs:
\begin{equation} \label{20}
Z' = \lambda_q Z + X, 
\end{equation}
where $\lambda_q$ controls the strength of the vector injection. The resulting combined representation will serve as an enhanced guide for the DMs. Intuitively, for sparse sequences with insufficient interactions, the closest encoding provides additional information that is most aligned with the user's interests. For noisy sequences, the extracted encoding helps amplify identifiable patterns and reduce the influence of irrelevant noise, thereby enhancing the expressive power of the guide.

In order to further improve the load balancing of experts, we introduce an auxiliary loss term to encourage different experts to receive balanced routing probabilities:
\begin{equation}
\mathcal{L}_{\text{balance}} = \sum_{k=1}^{K} \bar{p}_k \log (\bar{p}_k + \epsilon),
\label{21} %
\end{equation}
where, $\bar{p}_k = \frac{1}{N} \sum_{i=1}^{N} p_{i,k}$ represents the average allocation probability of the $k$-th expert. 
This hybrid expert mechanism not only retains the expressive power of multi-interest modeling but also achieves an end-to-end joint optimization process through differentiable routing and soft sparsity control.

\section{Experiment}
\begin{table*}[t]
\centering
\small  
\setlength{\tabcolsep}{1pt}  
\begin{tabular}{ll|ccc|cc|cccccc>{\bfseries}c}  
\toprule
Dataset & Metric & GRU4Rec & SASRec  & BERT4Rec & ComiRec & TiMiRec & DuoRec & CL4SRec & ACVAE & DreamRec & DiffuRec & DiQDiff & MGDiff \\  
\midrule
\multirow{6}{*}{ML-1M} 
& H@5   & 5.09 & 9.41 & 13.61 & 6.14 & 16.19  & 13.67 & 12.64 & 12.75 & 16.02 & 16.05 & 16.54 & 17.81 \\
& H@10  & 10.14 & 16.86  & 20.60 & 12.01 & 23.74  & 21.44 & 20.14 & 19.96 & 24.60 & 24.31 & 25.06 & 25.79 \\
& H@20  & 18.73 & 28.35  & 29.92 & 21.04 & 33.20  & 32.94 & 31.94 & 29.00 & 35.81 & 35.60 & 36.10 & 36.53 \\
& N@5   & 3.02 & 5.35  & 8.86 & 3.49 & 10.91  & 7.95 & 7.55 & 8.26 & 10.58 & 10.44 & 10.89 & 12.20  \\
& N@10  & 4.71 & 7.70  & 11.16 & 5.44 & 13.28  & 10.56 & 10.05 & 10.57 & 13.33 & 13.08 & 13.62 & 14.78 \\
& N@20  & 6.79 & 10.62  & 13.45 & 7.62 & 15.73  & 13.53 & 13.00 & 12.85 & 16.17 & 15.93 & 16.43 & 17.48 \\
\midrule
\multirow{6}{*}{Beauty} 
& H@5   & 0.99 & 3.24  & 2.16 & 2.02 & 1.93  & 5.34 & 5.28 & 2.50 & 5.28 & 5.36 & 5.37 & 6.13 \\
& H@10  & 1.97 & 6.29  & 3.69 & 4.42 & 3.37  & 7.60 & 7.32 & 3.91 & 7.10 & 7.24 & 7.54 & 8.40 \\
& H@20  & 3.82 & 9.01  & 5.82 & 7.67 & 5.20  & 10.69 & 10.61 & 6.09 & 10.59 & 10.54 & 10.89 & 11.35 \\
& N@5   & 0.64 & 2.37  & 1.29 & 1.08 & 1.21 & 3.31 & 3.00 & 1.72 & 3.25 & 3.85 & 4.04 & 4.41 \\
& N@10  & 0.87 & 3.26  & 1.86 & 1.80 & 1.73 & 4.22 & 4.02 & 2.17 & 4.16 & 4.46 & 4.58 & 5.14 \\
& N@20  & 1.41 & 3.69  & 2.32 & 2.62 & 2.19  & 5.02 & 4.88 & 2.73 & 4.96 & 5.37 & 5.35 & 5.88 \\
\midrule
\multirow{6}{*}{Toys} 
& H@5   & 1.13 & 4.50  & 1.96 & 2.27 & 1.19  & 5.63 & 5.51 & 2.22 & 5.44 & 5.61 & 5.76 & 6.45\\
& H@10  & 1.82 & 6.58  & 2.90 & 4.26 & 1.85  & 7.13 & 6.90 & 3.10 & 7.22 & 7.25 & 7.50 & 8.32\\
& H@20  & 3.21 & 9.20  & 4.62 & 6.91 & 2.75  & 9.84 & 10.12 & 4.44 & 9.65 & 9.87 & 9.94 & 10.86 \\
& N@5   & 0.67 & 3.04  & 1.13 & 1.19 & 0.74  & 3.08 & 3.37 & 1.59 & 4.01 & 4.21 & 4.33 & 4.85\\
& N@10  & 0.97 & 3.72  & 1.52 & 1.77 & 0.94  & 3.95 & 4.30 & 1.88 & 4.59 & 4.78 & 4.89 & 5.45\\
& N@20  & 1.24 & 4.36  & 1.87 & 2.43 & 1.17  & 4.74 & 5.11 & 2.21 & 5.19 & 5.38 & 5.51 & 6.09\\
\midrule
\multirow{6}{*}{Steam} 
& H@5   & 3.04 & 4.71 & 4.77 & 2.32 & 5.99 & 5.72 & 5.59 & 5.61 & 5.93 & 6.75 & 7.04 & 7.34\\
& H@10  & 5.40 & 8.41 & 7.91 & 5.41 & 9.70 & 9.75 & 9.48 & 9.31 & 9.65 & 10.54 & 11.32 & 11.60\\
& H@20  & 9.26 & 13.58 & 12.76 & 10.40 & 14.86 & 15.58 & 15.09 & 14.51 & 15.05 & 16.12 & 17.43& 17.61\\
& N@5   & 1.80 & 2.91 & 2.94 & 1.13 & 3.90 & 3.39 & 3.45 & 3.57 & 3.81 & 4.22 & 4.55 &4.84\\
& N@10  & 2.63 & 4.02 & 4.03 & 2.08 & 5.07 & 4.71 & 4.68 & 4.76 & 5.00 & 5.53 & 5.88&6.17\\
& N@20  & 3.53 & 5.39 & 5.17 & 3.37 & 6.39 & 6.17 & 6.09 & 6.07 & 6.36 & 7.14 & 7.43&7.68\\
% \midrule
% \multirow{6}{*}{Steam} 
% & H@5   & 3.04 & 4.71 & 4.77 & 2.32 & 5.99 & 5.72 & 5.59 & 5.61 & 5.93 & 6.75 & 7.04 &7.34\\
% & H@10  & 5.40 & 8.41 & 7.91 & 5.41 & 9.70 & 9.75 & 9.48 & 9.31 & 9.65 & 10.54 & 11.32&11.60\\
% & H@20  & 9.26 & 13.58 & 12.76 & 10.40 & 14.86 & 15.58 & 15.09 & 14.51 & 15.05 & 16.12 & 17.43& 17.61\\
% & N@5   & 1.80 & 2.91 & 2.94 & 1.13 & 3.90 & 3.39 & 3.45 & 3.57 & 3.81 & 4.22 & 4.55 &4.84\\
% & N@10  & 2.63 & 4.02 & 4.03 & 2.08 & 5.07 & 4.71 & 4.68 & 4.76 & 5.00 & 5.53 & 5.88&6.17\\
% & N@20  & 3.53 & 5.39 & 5.17 & 3.37 & 6.39 & 6.17 & 6.09 & 6.07 & 6.36 & 7.14 & 7.43&7.68\\
\bottomrule
\end{tabular}
\caption{Performance comparison of different recommendation models.}
\label{tab:model-performance}
\end{table*}
To present the experimental results addressing the following Research Questions (RQ), we have:
\begin{enumerate}
\item RQ1: How does our model perform compared to state-of-the-art models?
\item RQ2: What improvements do the designs of DSG and PAG bring to our model?
\item RQ3: Can the diversity of recommendations meet the diverse needs of users?
\item RQ4: How do hyper-parameters affect the performance of our model?

\end{enumerate}
\subsection{Experimental Settings}
We use four real - world datasets to validate the effectiveness of our model. All these datasets have been widely used for sequence recommendation:
\begin{itemize}
    \item The Amazon Beauty and Amazon Toys datasets consist of user reviews for beauty products and toys collected from the Amazon platform over nearly 20 years. 
    \item MovieLens - 1M is a widely - used benchmark dataset that contains 1 million movie ratings from 6,000 users on 4,000 movies. 
    \item The Steam dataset gathers information about video games available on the Steam platform, encompassing users’ playing time, prices, categories, and more.
\end{itemize}

% Following common data preprocessing methods~\cite{GRU4Rec:2,Diffrec:10,DiQDif:1,BERT4Rec:6},  we adopt a leave - one - out strategy for performance evaluation. Specifically, for all datasets, given a sequence \( \mathcal{S}=\{i_1, i_2, \ldots, i_n\} \), we use the most recent interaction (\(i_n\)) for testing, the penultimate interaction (\(i_{n - 1}\)) for model validation, and the earlier interactions (\(\{i_1, i_2, \ldots, i_{n - 2}\}\)) for model training. The length is set to 50 for the three datasets. The statistics of these datasets are reported in Table.~\ref{tab:datasets}. We can find that the average sequence lengths and scales of these datasets are very different, covering a wide range of real - world scenarios~\cite{Diffrec:10,DiQDif:1}. 

\subsubsection{Baseline}
MGDiff was evaluated against a range of leading methods in sequence recommendation, including traditional recommendation models, interest learning methods, contrastive learning approaches, and generative recommendation models, as detailed below:

\begin{itemize} \item \textbf{Traditional recommendation models}: GRU4Rec~\cite{GRU4Rec:2}, the first work to apply Gated Recurrent Units to sequence recommendation; SASRec~\cite{SASRec:3}, a widely‑adopted Transformer‑based model that captures long‑range dependencies through unidirectional self‑attention; and BERT4Rec~\cite{BERT4Rec:6}, which employs a bidirectional Transformer with masked item prediction to model the full context of user behavior sequences.
\item \textbf{Multi‑interest representation models}: ComiRec~\cite{ComiRe:7}, which extracts diverse user interests via dynamic routing and attention mechanisms; and its advanced successor TiMiRec~\cite{TimeRec:8}, which enhances interest discrimination through explicit target‑item supervision.
\item \textbf{Generative and contrastive learning models}: ACVAE~\cite{ACVAE:11}, which integrates adversarial training with contrastive learning within a variational autoencoder framework; CL4SRec~\cite{CL4SRec:12}, a contrastive learning framework that leverages sequence augmentations and consistency regularization; the pioneering diffusion‑based recommenders DreamRec~\cite{DreamRec:9} and DiffuRec~\cite{Diffrec:10}, which adopt denoising diffusion probabilistic models for next‑item generation; and the recent DiQDiff~\cite{DiQDif:1}, which introduces a quantile‑guided diffusion mechanism optimized via contrastive divergence maximization. \end{itemize}

\subsection{Overall Performance (RQ1)}
To answer  RQ1, we conducted a comprehensive comparative analysis on three real-world datasets. The overall performance comparison with state-of-the-art baselines is reported in Table~\ref{tab:model-performance}. Analyzing the results reveals a clear performance hierarchy driven by model evolution. First, traditional sequential models based on unidirectional or bidirectional Transformers (e.g., GRU4Rec, SASRec, and BERT4Rec~\cite{GRU4Rec:2,SASRec:3,BERT4Rec:6}) generally underperform. Their limitation lies in mapping users to single fixed vectors, which struggles to capture complex, multi-faceted user intents. Second, interest-based methods like ComiRec and TiMiRec~\cite{ComiRe:7,TimeRec:8} show improved performance by modeling diverse user interests, yet they are still bounded by the expressiveness of deterministic representations.
Notably, diffusion-based generative models (DreamRec, DiffuRec, and DiQDiff~\cite{DreamRec:9,DiQDif:1,Diffrec:10}) achieve the strongest performance among baselines. This superiority highlights the effectiveness of the generative paradigm, which models the conditional probability distribution of items rather than relying on discriminative point estimation, allowing for better handling of uncertainty in user behavior.
Most importantly, our proposed MGDiff consistently outperforms all baselines across all datasets, establishing a new state-of-the-art. Specifically, MGDiff yields substantial improvements, elevating HR@5 and NDCG@5 by approximately 8.6\% and 10\%, respectively, compared to the best performing baseline. This significant margin validates that our proposed Weight-adaptive Masking GNN and Popularity-Aware Guidance strategies effectively mitigate the semantic distortion and bias issues that limit existing diffusion models, leading to more precise and robust recommendations.

\subsection{Ablation Study (RQ2)}
\begin{table}[h]
\renewcommand\arraystretch{1} 
\small  
\setlength{\tabcolsep}{4pt}  
\begin{tabular}{l l c c c c c c} 
\toprule
Dataset & Ablation & H@5 & N@5 & H@10 & N@10 & H@20  & N@20 \\
\midrule
\multirow{4}{*}{Beauty} 
& Base    & 5.45    &   3.65  &  7.62   &  4.66   &   10.48  & 5.52  \\
& w/o DSG &  5.77   &  4.10   &  8.18   &  4.64   &   10.88  &  5.53   \\
& w/o PAG &  6.01   &  4.31   &  8.16   &  5.08   &   11.01  &  5.79   \\
& \textbf{MGDiff} &   \textbf{6.13}  &    \textbf{4.41} & \textbf{8.40}    &   \textbf{5.14}  &  \textbf{11.35}   &  \textbf{5.88}   \\
\midrule
\multirow{4}{*}{Toys} 
& Base    &  5.69   &  4.2   &  7.4   &  4.76   &   9.80  &  5.36   \\
& w/o DSG & 5.88    &   4.38   &   7.79   & 4.99    &  10.22   & 5.60    \\
& w/o PAG &   6.30  &  4.69   & 8.20    &  5.38   &   10.79  &  6.00   \\
& \textbf{MGDiff} &   \textbf{6.45}  &  \textbf{4.85}   & \textbf{8.32}    &   \textbf{5.45}  &  \textbf{10.86}   &  \textbf{6.09}   \\
\midrule
\multirow{4}{*}{ML-1M} 
& Base    & 15.22    &  10.05   &  23.50   & 12.71    & 35.09    &  15.62   \\
& w/o DSG & 15.30    &  10.25   &  23.82   & 12.98    & 34.98    &  15.79     \\
& w/o PAG & 17.07 & 11.69 & 25.32 & 14.33 & 36.34 & 17.10 \\
& \textbf{MGDiff} &  \textbf{17.81}   & \textbf{12.20}    &  \textbf{25.79}   &   \textbf{14.78}  &   \textbf{36.53}  &  \textbf{17.48}   \\
\midrule
\multirow{4}{*}{Steam} 
& Base    & 6.68    & 4.33 & 10.77   & 5.65    & 16.53  & 7.19  \\
& w/o DSG & 6.85    & 4.49   & 10.81   & 5.75    & 16.88   & 7.32    \\
& w/o PAG & 7.28   & 4.78  & 11.46  & 6.12 & 17.45   & 7.62  \\
& \textbf{MGDiff} &  \textbf{7.34}   & \textbf{4.84}    &  \textbf{11.60}   & \textbf{6.17}  & \textbf{17.61}  & \textbf{7.68}   \\
\midrule
\bottomrule
\end{tabular}
\caption{Ablation Experiments}
\label{tab:ablation_template}
\end{table}

To answer RQ2, we conducted an ablation study to verify the importance of our DSG and PAG modules. The experimental results are presented in Table.~\ref{tab:ablation_template}, where \textit{Base} refers to the variant excluding both modules, while \textit{w/o DSG} and \textit{w/o PAG} represent the variants of our model that omit the DSG module and the PAG module, respectively.

\begin{figure}[h]
    \centering
    % 第一行子图
    \begin{subfigure}[b]{0.235\textwidth}  % 宽度设为页面的45%（两个子图占满一行）
        \centering
        \includegraphics[width=\textwidth]{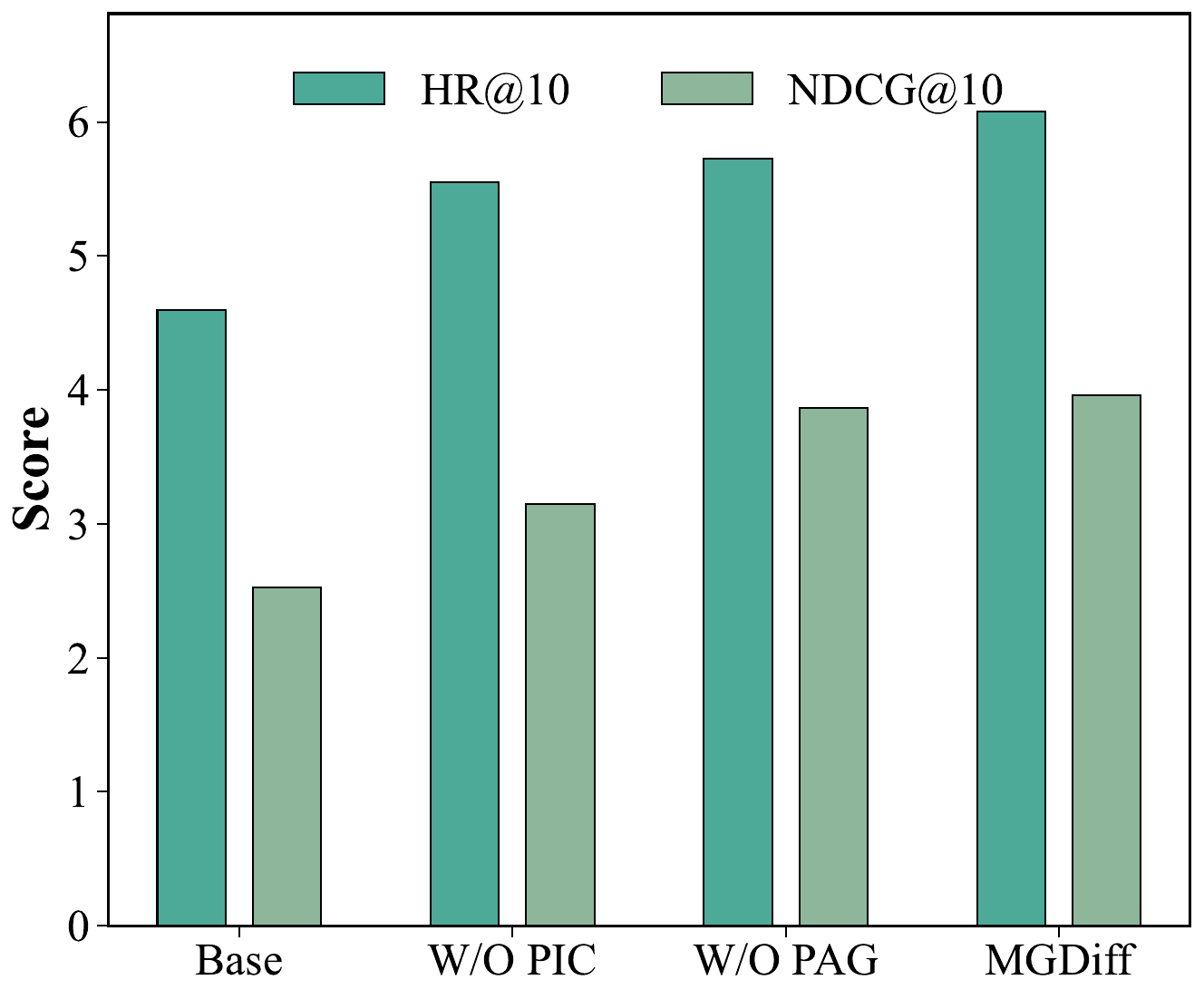}
        \caption{Cold Evaluation}
    \end{subfigure}
    \hfill  % 自动填充水平间隙，使子图右对齐
    \begin{subfigure}[b]{0.235\textwidth}
        \centering
        \includegraphics[width=\textwidth]{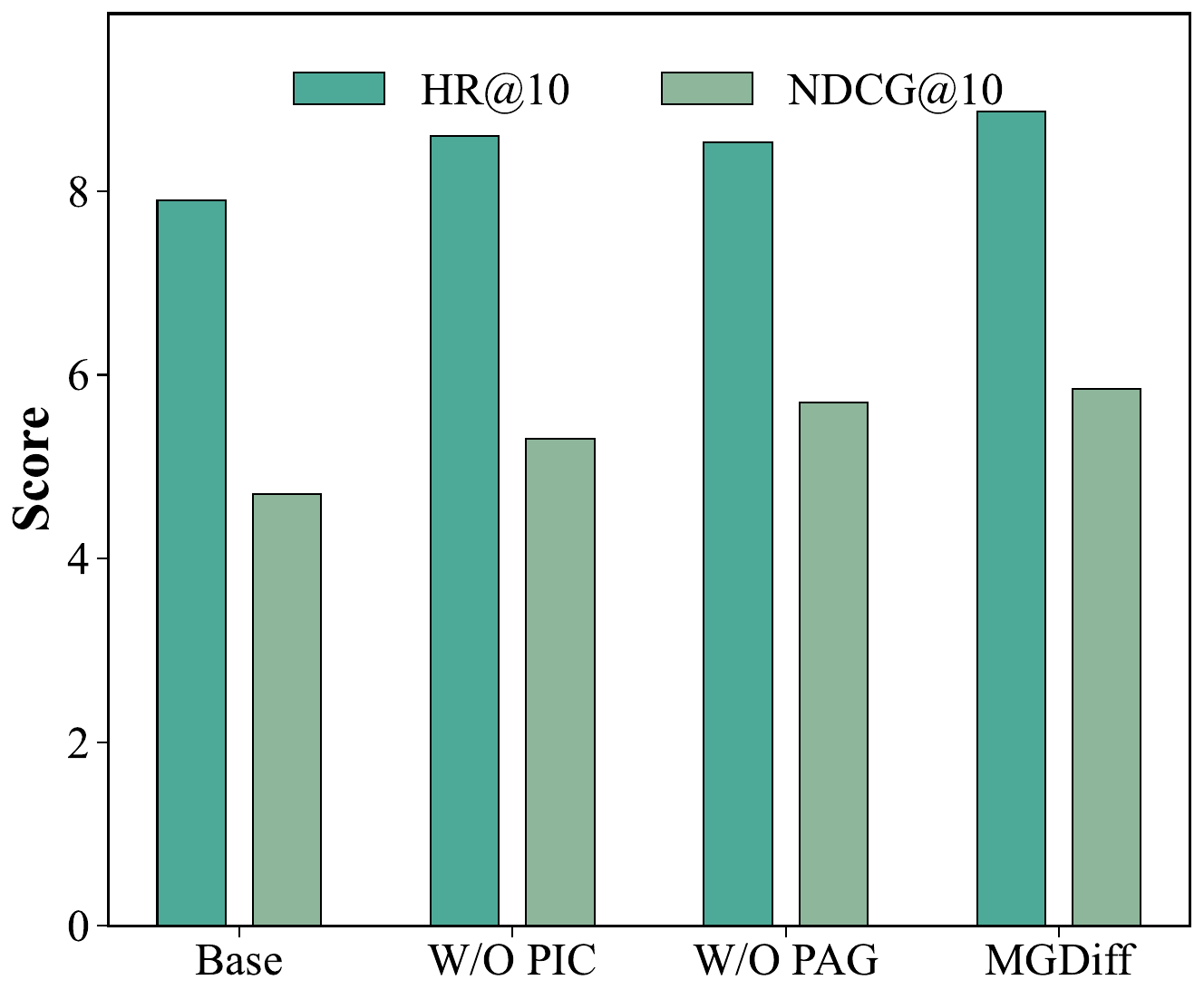}
        \caption{Short Evaluation}
    \end{subfigure}
    
    \caption{Experimental analysis comparing scenarios with little interaction history and those targeting long-tail items.}  % 总标题
    \label{fig:6}
\end{figure}

We observed that on all datasets, the performance of our \textit{w/o DSG} and \textit{w/o PAG} variants consistently outperforms that of \textit{Base}, which demonstrates the feasibility of each individual design. In addition, MGDiff showed the highest performance among the three variants, indicating that combining these two components further improved overall performance, demonstrating that they are stackable configurations.
Furthermore, we investigate the model's capability in handling data sparsity and popularity bias—the two core motivations of this work. We conducted a fine-grained analysis on the \textit{Toys} dataset, segmenting the data to isolate Sparse Users (users with the shortest 20\% interaction sequences) and Cold Items (items in the bottom 20\% popularity percentile). As shown in Fig.~\ref{fig:6}, our approach consistently outperforms baselines in these hard subsets. This highlights MGDiff's superiority in capturing user intent from scarce data and its ability to recommend diverse, long-tail items that traditional models often neglect.

\subsection{Visualization (RQ3)}
% To further validate MGDiff is capability to satisfy diverse user needs, we present t-SNE visualizations of project embeddings from different samples in Fig.~\ref{fig:5}. It can be observed that the distributions generated by the Base model and the \textit{w/o PAG} model are highly clustered, indicating a lack of diversity. In contrast, the \textit{w/o DSG} model and MGDiff yield more balanced and dispersed embedding distributions. This comparison reveals that the PAG component effectively alleviates the issue of niche products being overlooked during generation, thereby better accommodating varied user preferences. 需要恢复

Notably, while both \textit{w/o DSG} and MGDiff produce more diverse embeddings, certain clusters remain—which we consider reasonable, as real-world user interests often naturally converge. Furthermore, by comparing MGDiff with \textit{w/o DSG} and Base with \textit{w/o PAG}, we observe that incorporating structured guidance during generation further enhances the diversity of the embeddings produced by the diffusion model. This suggests that  DSG help in steering the model toward generating a richer variety of content.

% \begin{figure}[h]
%     \centering
%     % 第一行子图
%     \begin{subfigure}[b]{0.225\textwidth}  % 宽度设为页面的45%（两个子图占满一行）
%         \centering
%         \includegraphics[width=\textwidth]{Figure/T-SE2/kong.png}
%         \caption{Base}
%     \end{subfigure}
%     \hfill  % 自动填充水平间隙，使子图右对齐
%     \begin{subfigure}[b]{0.225\textwidth}
%         \centering
%         \includegraphics[width=\textwidth]{Figure/T-SE2/cdm.png}
%         \caption{w/o DSG}
%     \end{subfigure}
%     \\  % 强制换行，开始第二行
%     % 第二行子图
%     \begin{subfigure}[b]{0.225\textwidth}
%         \centering
%         \includegraphics[width=\textwidth]{Figure/T-SE2/svq.png}
%         \caption{w/o PAG}
%     \end{subfigure}
%     \hfill
%     \begin{subfigure}[b]{0.225\textwidth}
%         \centering
%         \includegraphics[width=\textwidth]{Figure/T-SE2/quankai.png}
%         \caption{MGDiff}
%     \end{subfigure}
%     \caption{The T-SNE visualization of the generated item embeddings on the Beauty dataset.}  % 总标题
%     \label{fig:5}
% \end{figure}

\begin{table}
\centering
\small
\setlength{\tabcolsep}{2.3pt}
\begin{tabular}{llccccc}
\toprule
Dataset & Model & ARP $\downarrow$ & Coverage $\uparrow$ & Gini $\downarrow$ & Tail@20 $\uparrow$ & H@20$\uparrow$  \\
\midrule
\multirow{6}{*}{Toys}
& DiffRec & 2.977 & 0.947 & 0.653 & 0.053 &6.20 \\
& CL4SRec & 3.245 & 0.827 & 0.778 & 0.035 & 3.53\\
& DiQdiff & 2.958 & 0.944 & 0.664 & 0.056 & 6.83\\
& Base & 3.148 & 0.900 & 0.719 & 0.0351 & 5.93\\
& w/o DSG & \textbf{2.823} & 0.975 & 0.625 & 0.068 & 7.19 \\
& w/o PAG & 2.888 & 0.965 & 0.629 & 0.065 & 7.32\\
& MGDiff & 2.837 & \textbf{0.989} & \textbf{0.603} & \textbf{0.068} &\textbf{7.55}\\
\midrule
\multirow{6}{*}{Beauty}
& DiffRec & 3.319 & 0.933 & 0.713 & 0.038 &3.52 \\
& CL4SRec & 3.916 & 0.725 & 0.867 & 0.022 & 3.53\\
& DiQdiff & 3.294 & 0.943 & 0.704 & 0.042 &  3.79\\
& Base & 3.451 & 0.911 & 0.749 & 0.027 & 3.53\\
& w/o DSG & 3.267 & 0.952 & 0.687 & 0.046 &3.81 \\
& w/o PAG & 3.240 & 0.967 & 0.684 & 0.046 & 4.21 \\
& MGDiff & \textbf{3.214} & \textbf{0.975} & \textbf{0.660} & \textbf{0.047} &\textbf{4.41} \\
\bottomrule
\end{tabular}
\caption{Debiasing Performance Comparison on Toys and Beauty Datasets}
\label{tab:debias_metrics}
\end{table}

\subsection{Long-tail Performance Analysis}
To further validate the model’s effectiveness in eliminating popularity bias, we compared MGDiff with traditional recommendation models and diffusion-based methods on the Toys and Beauty datasets. The experimental results are shown in Table.~\ref{tab:debias_metrics}. Compared to other methods, MGDiff achieved lower ARP and Gini coefficients on both datasets, indicating that it effectively mitigates the tendency to over-recommend popular items. At the same time, MGDiff significantly improved coverage and the Tail@20 metric, reflecting its ability to broaden the scope of item recommendations and increase exposure opportunities for long-tail items. Furthermore, we focused on users whose interactions primarily revolve around long-tail items and conducted a further analysis of the H@20 metric for this user group. The results show that MGDiff also performs best in this scenario, indicating that our method not only increases the exposure of long-tail items but also effectively improves the recommendation accuracy for users with niche preferences. The quantitative experiments described above consistently demonstrate that MGDiff can mitigate popularity bias while comprehensively optimizing recommendation quality, making it particularly suitable for long-tail recommendation scenarios.

% \begin{figure}[htbp] 
%     \centering
%     \includegraphics[width=0.5\textwidth]{Figure/Hyper-parameter Analysis.png}
%     \caption{Sensitivity of model to the hyperparameter $\lambda_q$ and $\lambda_c$.}
%     \label{fig:7}
% \end{figure}

% \begin{figure}[h]
%     \centering
%     % 第一行子图
%     \begin{subfigure}[b]{0.235\textwidth}  % 宽度设为页面的45%（两个子图占满一行）
%         \centering
%         \includegraphics[width=\textwidth]{Figure/Hyperparamter/beauty_hr5_times_centered.png}
%         \caption{Beauty HR@5}
%     \end{subfigure}
%     \hfill  % 自动填充水平间隙，使子图右对齐
%     \begin{subfigure}[b]{0.235\textwidth}
%         \centering
%         \includegraphics[width=\textwidth]{Figure/Hyperparamter/ML-1M_hr5_times_centered.png}
%         \caption{ML-1M HR@5}
%     \end{subfigure}
%      \hfill  % 自动填充水平间隙，使子图右对齐
%     \begin{subfigure}[b]{0.235\textwidth}
%         \centering
%         \includegraphics[width=\textwidth]{Figure/Hyperparamter/toys_hr5_times_centered.png}
%         \caption{Toy HR@5}
%     \end{subfigure}
%      \hfill  % 自动填充水平间隙，使子图右对齐
%     \begin{subfigure}[b]{0.235\textwidth}
%         \centering
%         \includegraphics[width=\textwidth]{Figure/Hyperparamter/Steam_hr5_times_centered.png}
%         \caption{Steam HR@5}
%     \end{subfigure}
    
%     \caption{Sensitivity of model to the hyperparameter $\lambda_q$ and $\lambda_c$.}  % 总标题
%     \label{hyfig:7}
% \end{figure}

\begin{figure}[h]
    \centering
    % 第一行子图
    \begin{subfigure}[b]{0.235\textwidth}  % 宽度设为页面的45%（两个子图占满一行）
        \centering
        \includegraphics[width=\textwidth]{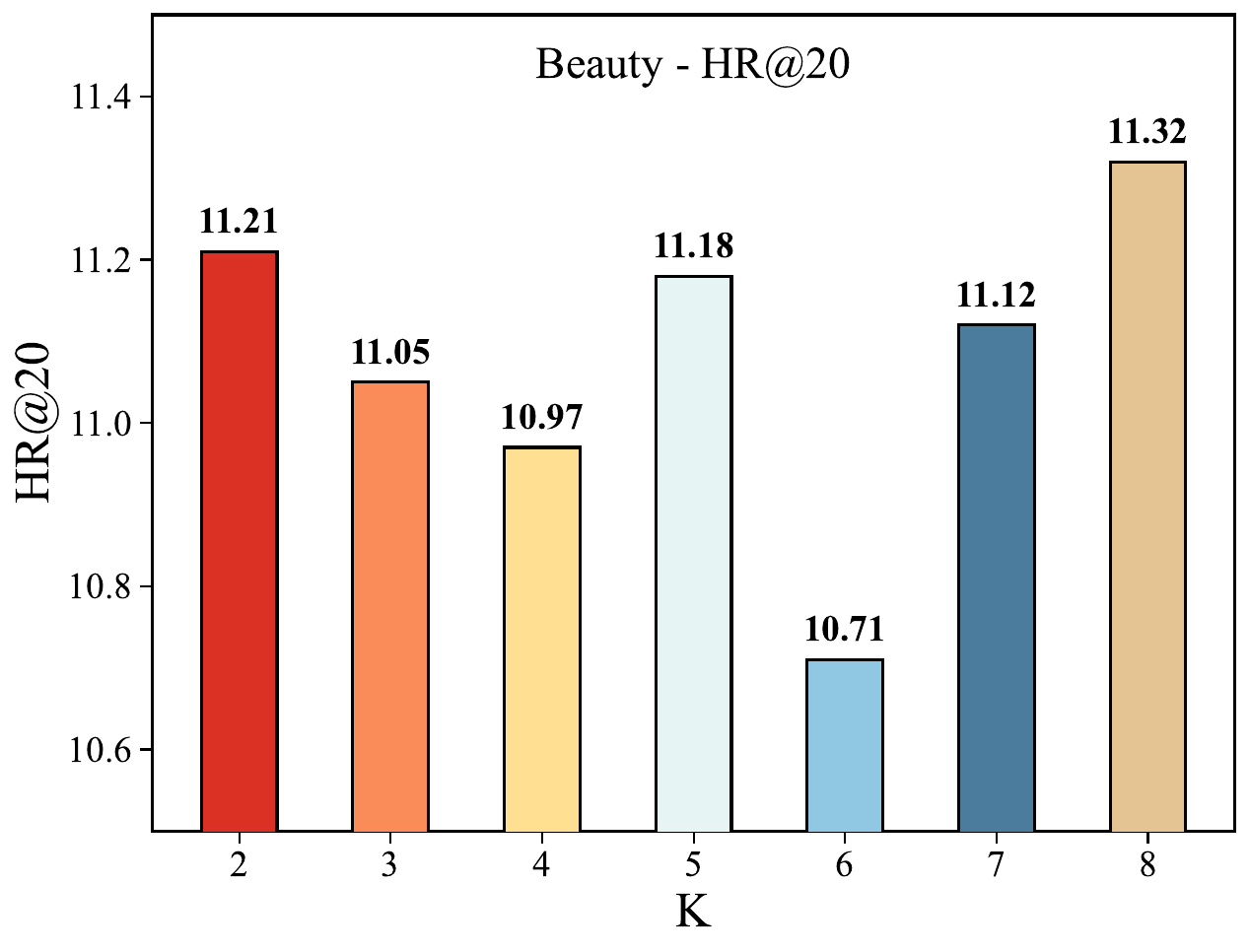}
        \caption{Beauty HR@20}
    \end{subfigure}
    \hfill  % 自动填充水平间隙，使子图右对齐
    \begin{subfigure}[b]{0.235\textwidth}
        \centering
        \includegraphics[width=\textwidth]{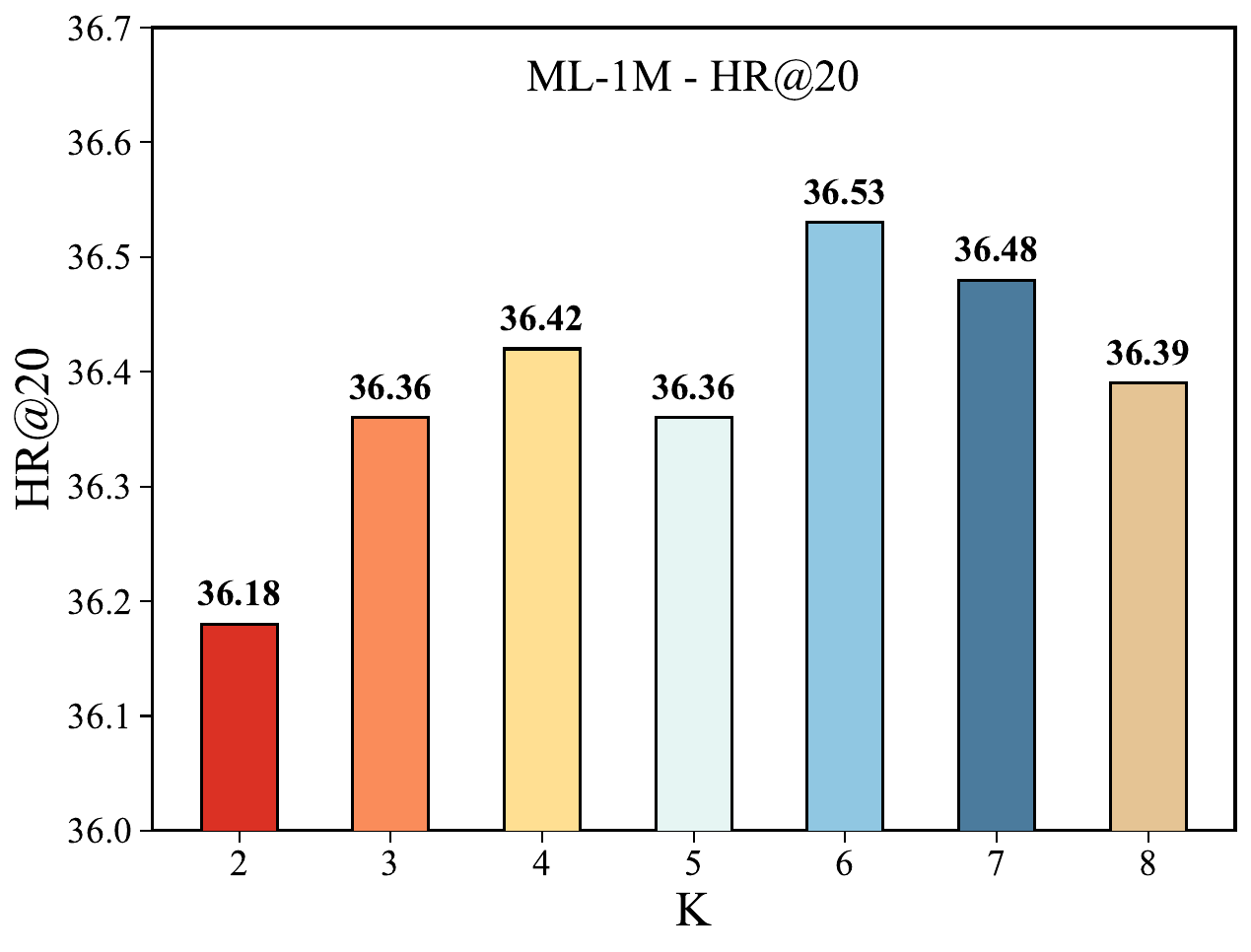}
        \caption{ML-1M HR@20}
    \end{subfigure}
     \hfill  % 自动填充水平间隙，使子图右对齐
    \begin{subfigure}[b]{0.235\textwidth}
        \centering
        \includegraphics[width=\textwidth]{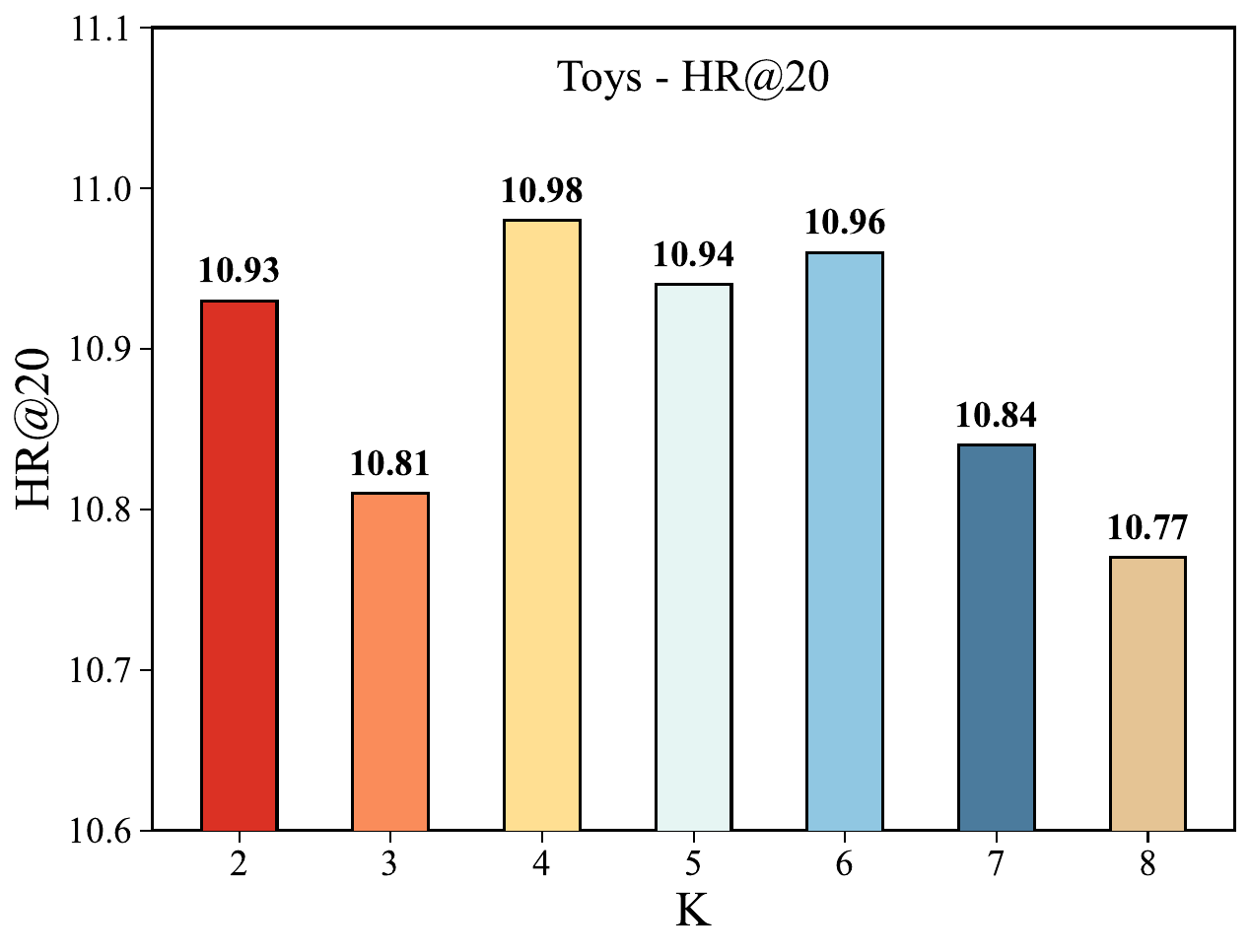}
        \caption{Toy HR@20}
    \end{subfigure}
     \hfill  % 自动填充水平间隙，使子图右对齐
    \begin{subfigure}[b]{0.235\textwidth}
        \centering
        \includegraphics[width=\textwidth]{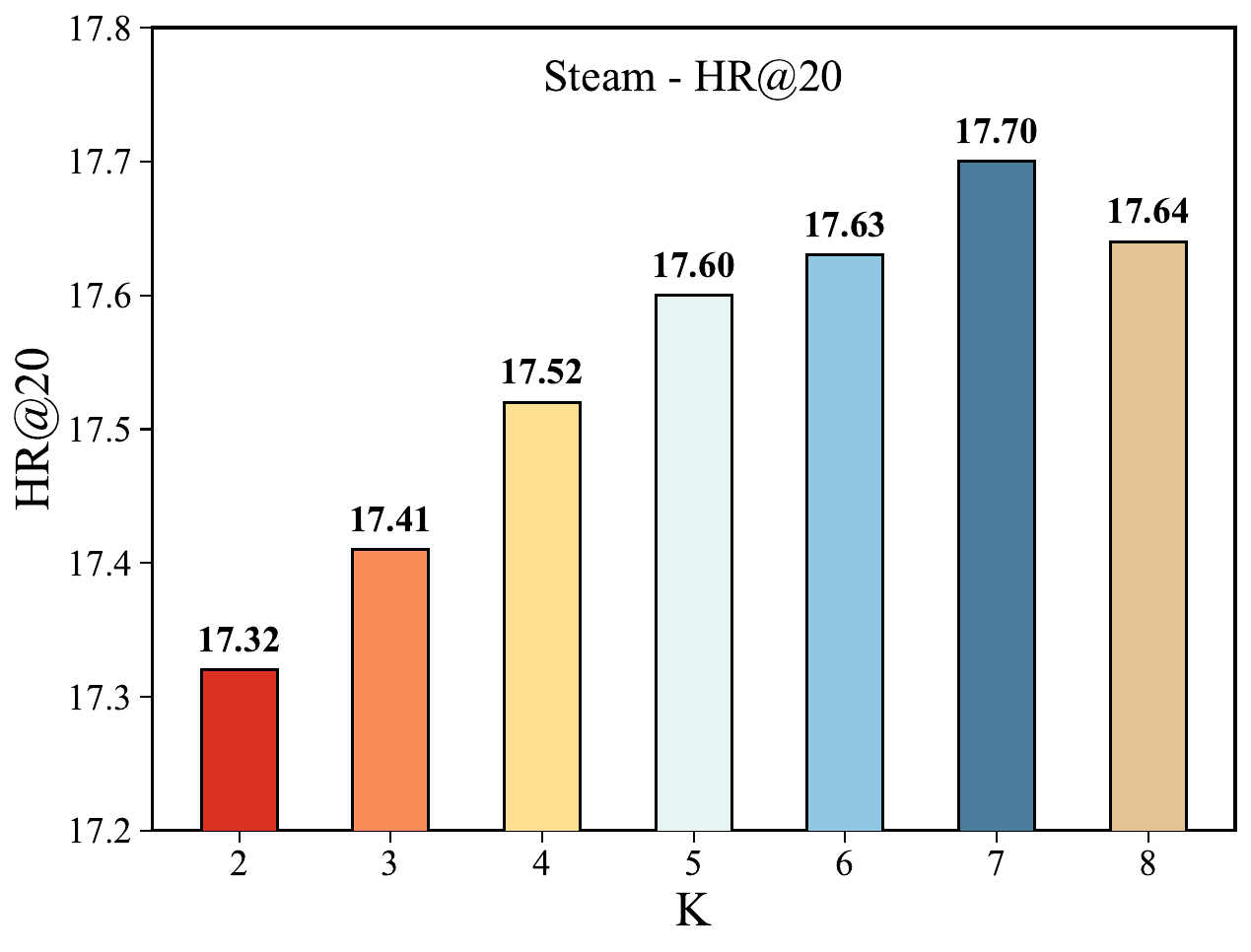}
        \caption{Steam HR@20}
    \end{subfigure}
    
    \caption{Performance analysis with different numbers of experts.}  % 总标题
    \label{fig:expert}
\end{figure}

\section{Conclusion}
In this paper, we propose MGDiff, a novel framework designed to resolve the critical bottlenecks of semantic distortion and endogenous generation bias in diffusion-based sequence recommendation. To address the semantic distortion caused by the sparsity and randomness of interaction data, we devise a Dual-layer Semantic Guidance framework. This component synergizes a Weight-adaptive Masking GNN, which leverages curriculum learning to uncover latent item correlations—with a dynamically routed Mixture-of-Experts network that adaptively refines user interest modeling from broad exploration to precise specialization. Furthermore, to counteract the Matthew effect and mode collapse caused by popularity bias, we introduce a Popularity-Aware Guidance  strategy. By incorporating popularity as a differentiable adjustment signal within a contrastive learning objective, PAG effectively enhances item discriminability, ensuring a robust balance between recommendation accuracy and diversity. Extensive experiments on real-world datasets demonstrate that MGDiff consistently outperforms state-of-the-art baselines.

% \section{Acknowledge}
% This work was supported in part by the National Natural Science Foundation of China (Nos. 62502101, 62462002) and partially supported by the Natural Science Foundation of Guangxi, China (Nos. 2025GXNSFAA069958, 2025GXNSFBA069394).

\bibliographystyle{IEEEtran}
\bibliography{my}
\end{document}